\documentstyle[preprint,eqsecnum,aps]{revtex}

\begin{document}
\draft
\preprint{HEP/123-qed}
\title{
Anisotropic charge transfer mechanism in La$_{2-x}$Sr$_x$CuO$_4$ and 
Bi$_2$Sr$_2$Ca$_{1-x}$Y$_x$Cu$_2$O$_{8+\delta}$
}

\author{S. Sugai, N. Hayamizu, and T. Hosokawa
}
\address{Department of Physics, Faculty of Science, 
Nagoya University, Chikusa-ku, Nagoya 464-8602, Japan} 

\date{\today}
\maketitle

\begin{abstract}
Raman spectra of La$_{2-x}$Sr$_x$CuO$_4$ are similar to those of 
Bi$_2$Sr$_2$CaCu$_2$O$_{8+\delta}$ except for the split two-magnon peaks 
at the stripe phase and the stronger step-like decrease in the low 
energy $B_{\rm 2g}$ spectra.  
The suppressed $B_{\rm 1g}$ spectral region below the two-magnon peak 
decreases as the carrier density increases, whereas the energy of 
the step-like suppression is little dependent on the carrier density.  
It suggests the anisotropic charge transfer mechanism coupled with 
magnetic excitations along the nearest neighbor Cu sites and the phononic 
excitations along the diagonal direction.  
\end{abstract}

\pacs{PACS numbers: 74.72.Dn, 74.72.Hs, 74.25.Gz}

\narrowtext
	Angle resolved photoemission spectroscopy (ARPES) disclosed the 
similarity  and the difference in La$_{2-x}$Sr$_x$CuO$_4$ (LSCO) and 
Bi$_2$Sr$_2$Ca$_1$Cu$_2$O$_{8+\delta}$ (BSCCO).  
In BSCCO the low energy spectra at $(\pi, 0)$ are suppressed at low 
carrier densities via the interaction with collective magnetic 
excitations at $(\pi, \pi)$ \cite{1,2,3,4}.  
The energy region of the suppressed intensity decreases as the carrier 
density increases.  
The spectra at $(\pi/2, \pi/2)$ are not suppressed at all carrier 
densities.  
On the other hand, in the spectra at $(\pi, 0)$ of LSCO a new hump 
appears at the insulator-metal transition 
and the hump develops to a peak with decreasing the energy as the 
carrier density increases \cite{5,6,7}.  
The low energy region at $(\pi/2, \pi/2)$ is strongly depleted at 
$x\le 0.12$ and suddenly increases at $x=0.15$.  
These anomalous features in LSCO were attributed to the stripe 
structure \cite{8,9,10}.  

	Wide-energy Raman scattering can observe electronic excitations 
at the selective $k$-space region along $(0, 0)-(\pi, 0)$, and 
$(0, 0)-(\pi, \pi)$ correspondingly to ARPES \cite{11,12}.  
In Bi$_2$Sr$_2$Ca$_{1-x}$Y$_x$Cu$_2$O$_{8+\delta}$ (BSCYCO) the Raman 
spectra are consistent with the results of ARPES \cite{13}, but in LSCO 
they are rather different from ARPES \cite{14,15,16}.  
The 300 K spectra, in which the stripe structure almost disappears at all 
carrier densities in LSCO, are qualitatively same in both superconductors 
except for the enhanced step-like decrease below about 1100 cm$^{-1}$ in 
the $B_{\rm 2g}$ spectra of LSCO.  
As temperature decreases, the spectra do not change so much except 
in the stripe phase of LSCO ($0.035\le x \le 0.06$ and $x=0.115$).  

	The Raman spectroscopy is valuable as an alternative 
method to measure the electronic excitation spectra to ARPES, because 
Raman spectra can detect anisotropic superconducting gap at $(\pi, 0)$ 
and $(\pi/2, \pi/2)$ in consistent with other 
experiments \cite{13,17,18,19,20} and the wide-energy Raman spectra in 
BSCYCO are consistent with the ARPES \cite{13}.  
The merit of Raman scattering is less sensitivity to the surface 
condition.  
The measurement of ARPES of LSCO is limited to low temperatures 
from the instability of the surface at high temperatures, while 
Raman scattering can measure up to about 500-600 K 
without disturbing the electronic states.   

	Raman scattering detects $B_{\rm 1g}$ and 
$B_{\rm 2g}$ excitations separately by choosing the polarizations 
of incident and scattered light as $(\hat{\mbox{\boldmath$E$}_i}, 
\hat{\mbox{\boldmath$E$}_s})=(xy)$ and $(ab)$, respectively \cite{11,12}.  
Here $a=[1, 0, 0]$ and $b=[0, 1, 0]$ are directions along the Cu-O-Cu 
bonds, and $x=[1, 1, 0]$ and $y=[1, -1, 0]$ are diagonal directions.  
The base function of $B_{\rm 1g}$ is $k_{x^2}-k_{y^2}$ and then it has 
a maximum at the $(0, 0)-(\pi, 0)$ direction.  
Thus the $B_{\rm 1g}$ electronic Raman spectrum represents the excitation 
along the $(0, 0)-(\pi, 0)$.  On the other hand the base function of 
$B_{\rm 2g}$ is $k_xk_y$ and has a maximum at $(0, 0)-(\pi, \pi)$.  
The $B_{\rm 2g}$ spectrum represents the excitation along 
$(0, 0)-(\pi, \pi)$.  

	The single crystals were synthesized by the TSFZ method 
utilizing an infrared radiation furnace with four elliptic mirrors 
(Crystal system, FZ-T-4000).  
The N\'eel temperature ($T_{\rm N}$) at $x=0$ is 293 K.  
The superconducting transition temperature $T_{\rm c}$'s determined 
from the middle point of the transition in the electric resistivity 
are 12 K $(x=0.06)$, 33 K $(x=0.1)$, 33 K $(x=0.115)$, 42 K 
$(x=0.15)$, 32 K $(x=0.2)$, and 13 K $(x=0.25)$.  
The details will be presented elsewhere \cite{21}.  

	Raman spectra were measured on fresh cleaved surfaces in a 
quasi-back scattering configuration utilizing a triple monochromator 
(JASCO, NR-1810), a liquid nitrogen cooled CCD detector (Princeton, 
1100PB), and a 5145 \AA \ Ar-ion laser (Spectra Physics, stabilite 
2017).  
The cleaved surfaces give intrinsic spectra which are different from 
those on the surfaces prepared by mechanical polishing or chemical 
etching \cite{21}.  
The laser beam of 10 mW was focused on the area of 50 
$\mu$m$\times$500 $\mu$m.  
The obtained Raman intensity was corrected by the spectroscopic 
efficiency of the optical system.

	Figure 1 shows the $B_{\rm 1g}$ and $B_{\rm 2g}$ Raman 
spectra at 100 K in the normal phases of optimally doped LSCO and 
BSCYCO.  
The solid curves indicate the expected ideal electronic Raman 
spectra in the strongly correlated metal.  
The $B_{\rm 1g}$ Raman spectra in overdoped LSCO and BSCYCO 
show these ideal metallic spectra.  
As the carrier density decreases the suppressed region 
increases at low energies.  
On the other hand the $B_{\rm 2g}$ spectra are very different 
from $B_{\rm 1g}$.  
The carrier density dependence is very small.  
The suppression in the low energy spectra is much larger in LSCO 
than in BSCYCO.  

	Figure 2 shows the carrier density dependence of the 
$B_{\rm 1g}$ and the $B_{\rm 2g}$ spectra of LSCO at 5 K and 300 K.  
The intensity is normalized so that the total integrated intensity 
below 6000 cm$^{-1}$ is the same.  
The sharp peaks below 700 cm$^{-1}$ are due to one-phonon 
scattering and the peaks from 700 to 1400 cm$^{-1}$ are due to 
two-phonon scattering.  
The two-phonon intensity is much larger in the $A_{\rm 1g}$ spectrum 
than in the $B_{\rm 2g}$ spectrum.  
The 3117 cm$^{-1}$ peak at 300 K in the $B_{\rm 1g}$ spectra of the 
insulating phase $(x=0)$ is the two-magnon peak.  
This energy is near $3J$, where $J$ is the exchange interaction 
energy between the nearest neighbor spins at copper sites.  
As the carrier density increases, the peak energy at 300 K shifts 
monotonically to low energy via electron-magnon interactions.  
It is consistent with the experiments by Naeini {\it et al.} \cite{14,15}.  
The peak energies are shown in Fig. 3.  
The $B_{\rm 1g}$ spectrum of overdoped sample $(x=0.25)$ is close 
to the ideal metallic spectrum in Fig. 1.  
The overall spectra at 5 K are similar to that at 300 K, unless 
stripe structure develops.  
However, in the stripe phase ($0.035 \le x \le 0.06$ and $x=0.115$) 
at 5 K, the two-magnon peak splits into double peaks near $3J$ and 
$2J$ \cite{16}.  
The clear split of the two-magnon peak is observed only at narrow 
carrier density region corresponding to the decrease of $T_{\rm c}$ 
measured by electric resistivity.  
It is very different from the results of neutron scattering in which 
the incommensurate magnetic peaks have been observed at whole carrier 
densities \cite{22,23,24}.         
It is also different from ARPES in which the carrier density 
dependence for the stripe phase is not observed.  
In the stripe phase, the low energy spectra below about 1100 cm$^{-1}$ 
are just like the metallic spectra \cite{16}.  
The threshold energy is exactly the same as one of the two-phonon 
energies in the $A_{\rm 1g}$ spectra.  
This phonon is assigned to the zone boundary $(\pi, 0, 0)$ longitudinal 
optical (LO) phonon mode with anomalous dispersion measured by neutron 
scattering \cite{25,26,27}.  
Oxygen atoms are vibrating in this mode.  

	The decrease of the two-magnon peak energy at 300 K, as the 
carrier density increases, can be considered as the decrease of the 
suppressed spectral region below the two-magnon peak.  
ARPES at $(\pi, 0)$ and 11 K showed the appearance of a new peak at about 
0.1 eV at $x=0.05$.  
The binding energy decreases and the intensity increases, as the 
carrier density increases \cite{7}.  
The energy is lower than a half of the $B_{\rm 1g}$ peak energy.  
The spectra do not change by the formation of the stripe 
structure at $0.035\le x \le 0.06$ and $x=0.115$ contrary to the 5 K 
Raman spectra in Fig. 2.  
In the case of BSCYCO the carrier density dependence is consistent 
with the ARPES.  

	As for the $B_{\rm 2g}$ spectra at 300 K, the carrier 
density dependence is very different.  
The 4170 cm$^{-1}$ peak at $x=0$ is due to the two-magnon scattering.  
In the $B_{\rm 2g}$ symmetry the two-magnon scattering is inactive, 
if only the nearest neighbor exchange interaction is considered \cite{28}.  
The diagonal next neighbor interaction and four spin cyclic exchange 
interaction can make it active \cite{28,29,30}.  
The peak energy 4170 cm$^{-1}$ is just $4J$, that is, twice the zone 
boundary magnon energy.  
It suggests that the magnon-magnon interaction does not work.  
The two-magnon peak energy decreases from $x=0$ to 0.1 and becomes 
constant above it as shown in Fig. 3.  
In the stripe phase at 
$0.035 \le x \le 0.06$ and $x=0.115$, the spectra change into the similar 
structure to the $B_{\rm 1g}$ spectra, as temperature decreases to 5 K, 
because the crystal symmetry relaxes by the stripe structure.  
The spectrum is composed of the split two-magnon peaks at about $3J$ 
and $2J$ above about 1100 cm$^{-1}$ and the metallic spectrum below 
it.  
This metallic spectrum can be viewed as a step-like decrease of the 
scattering intensity below about 1100 cm$^{-1}$.  This step-like 
decrease is commonly observed at 5 K even in the compound in which the 
stripe structure does not develop.  
And also this step-like decrease is reserved even at 300 K at which 
the stripe structure almost disappears even at $0.035 \le x \le 0.06$ 
and $x=0.115$.  
The energies of the valleys at the steps are shown in 
Fig. 3.  
The energies are almost independent of the carrier density.  
This suggests that the step-like decrease does not result from the 
magnetic origin, because the collective magnetic excitation energy 
decreases clearly with the increase of the carrier density as 
shown in the $B_{\rm 1g}$ two-magnon peak energy in Fig. 3.  
The origin is possibly the electron-phonon interaction with the zone 
boundary $(\pi, 0, 0)$ LO phonon, because the energy is the same as 
discussed in the $B_{\rm 1g}$ spectra of the stripe phase \cite{25,26,27}.  

	The $(\pi/2, \pi/2)$ ARPES spectra subtracted the (0, 0) spectra 
is very different from the $B_{\rm 2g}$ Raman spectra \cite{7}.  
The low-energy intensity is very weak at $x\le 0.12$.  
The carrier density dependence relating to the stripe phase 
($0.035\le x \le 0.06$ and $x=0.115$) is not observed.  
The sudden increase of the low energy spectral intensity at $x=0.15$ in 
the ARPES spectrum is inconsistent with the Raman spectrum.  

	The step-like decrease below about 1100 cm$^{-1}$ is 
also observed in lightly underdoped to overdoped BSCYCO \cite{13}.  
Figure 4 shows the $B_{\rm 2g}$ spectra at $x=0.3$ (underdoping 
$T_{\rm c}$=75 K), $x=0.2$ (underdoping $T_{\rm c}$=87 K), 
$x=0.1$ (optimal doping $T_{\rm c}$=95 K), and $x=0$ (overdoping, 
$T_{\rm c}$=84 K).  
The low energy peaks at 20 K are the superconducting peaks \cite{13}.  
The step-like decrease is not observed at $x=0.3$, but it increases as 
the carrier density increases and the temperature decreases 
from 300 K to 100 K.  The stripe structure is probably limited to much 
narrower carrier density region than in LSCO, even if it 
exists \cite{31}.  
Therefore the suppression of the low energy $B_{\rm 2g}$ spectra at wide 
carrier density region indicates again that the suppression in the 
$B_{\rm 2g}$ spectra is not related to the stripe structure.  
The stronger suppression in LSCO than in BSCYCO results from the stronger 
electron-phonon interaction in LSCO.  
It is consistent with the stronger two-phonon Raman intensity in LSCO than 
in BSCYCO.  

	The phononic origin for the depletion in the $B_{\rm 2g}$ spectra 
can be confirmed by comparing the energy of the step between oxide and 
sulfide, because almost only chalcogen atoms are vibrating in the relevant 
LO phonon mode.   
In the metallic phase of BaCo$_{1-x}$Ni$_x$S$_2$ (BCNS), the same small 
step-like decrease is observed 
below about 480 cm$^{-1}$ \cite{32}.  
The energy difference, 1100 cm$^{-1}$ in LSCO and 480 cm$^{-1}$ in BCNS, 
indicates that the origin of the step-like decrease in $B_{\rm 2g}$ is 
the electron-phonon interaction.  

	The present Raman scattering experiment clearly shows the 
different origin of the spectral suppression in the $B_{\rm 1g}$ spectra 
corresponding to the excitations along $(0, 0)-(\pi, 0)$ and the 
$B_{\rm 2g}$ spectra along $(0, 0)-(\pi, \pi)$.  
That is, the electronic excitations along $(0, 0)-(\pi, 0)$ is affected by 
the electron-magnon interaction, whereas those along $(0, 0)-(\pi, \pi)$ 
by the electron-phonon interaction.  
It can be interpreted in the real space that the charge transfer along 
the Cu-O-Cu direction is coupled with magnetic excitations, while the 
charge transfer along the diagonal direction is coupled with phonons.  
It is plausible, because the direction of spins are opposite at the 
nearest neighbor Cu sites but the same at the diagonal sites.  
The anomalous LO phonon with the carrier density dependent energy is 
observed at $(\pi, 0)$ \cite{25,26,27}.
The breathing phonon mode clusters formed from the LO phonon modes at 
$(\pi, 0)$ and $(0, \pi)$ are located along the diagonal direction.  
It is consistent with the above interpretation.  

	In conclusion, the wide-energy $B_{\rm 1g}$ and the $B_{\rm 2g}$ 
Raman spectra provide electronic excitation spectra corresponding to 
ARPES along $(0, 0)-(\pi, 0)$ and $(0, 0)-(\pi, \pi)$, respectively.  
In the $B_{\rm 1g}$ spectra the suppression below the two-magnon peak 
decreases as the carrier density increases, whereas in the 
$B_{\rm 2g}$ spectra the step-like suppression is little dependent on the 
carrier density.  
The former can be attributed to the results of the electron-magnon 
interaction and the latter to the electron-phonon interaction.  
The above results suggest that the charge transfer mechanism is 
anisotropic in the real space, that is, electrons couple dominantly with 
magnetic excitations for the transfer along the Cu-O-Cu direction 
and with phonons along the diagonal direction.  
These anisotropic charge transfer mechanisms are probably related to the 
high $T_{\rm c}$ superconducting mechanism.

Acknowledgments - 
This work was supported by CREST of the Japan Science and 
Technology Corporation.

\begin{figure}
\caption[]{
The $B_{\rm 1g}$ and $B_{\rm 2g}$ Raman spectra in the normal phase at the 
optimal doping of La$_{2-x}$Sr$_x$CuO$_4$ and 
Bi$_2$Sr$_2$Ca$_{1-x}$Y$_x$Cu$_2$O$_{8+\delta}$.  
The $B_{\rm 1g}$ spectra approach the typical metallic spectra 
(solid curves) as the carrier density increases.  
However, the $B_{\rm 2g}$ spectra approach the typical metallic spectra as 
the carrier density decreases in 
Bi$_2$Sr$_2$Ca$_{1-x}$Y$_x$Cu$_2$O$_{8+\delta}$.  
}
\label{fig1}
\end{figure}

\begin{figure}
\caption[]{
The carrier density dependence of the $B_{\rm 1g}$ and the 
$B_{\rm 2g}$ spectra at 300 K at which the stripe structure almost 
disappears and at 5 K at which the influence of the stripe is observed at 
$0.035\le x \le 0.06$ and $x=0.115$ by the split of the two-magnon peaks.  
The intensity is normalized so that the integrated intensity below 6000 
cm$^{-1}$ is the same.  
}
\label{fig2}
\end{figure}

\begin{figure}
\caption[]{
The carrier density dependence of the $B_{\rm 1g}$ and the 
$B_{\rm 2g}$ two-magnon peak energies at 300 K and the energies at the 
valley (or kink) in the $B_{\rm 2g}$ spectra at 5 K and 300 K.  
The small effect of the split two-magnon peak is retained in the 
$B_{\rm 2g}$ spectra at 300 K.  
}
\label{fig3}
\end{figure}

\begin{figure}
\caption[]{
The temperature and carrier density dependence of the $B_{\rm 2g}$ 
Raman spectra in Bi$_2$Sr$_2$Ca$_{1-x}$Y$_x$Cu$_2$O$_{8+\delta}$.  
The $T_{\rm c}$'s are shown in the parentheses.  
UD: underdoping, OP: optimal doping, and OD: overdoping.
}
\label{fig4}
\end{figure}

\end{document}